\documentclass{PoS}

\usepackage[utf8]{inputenc}
\usepackage{t1enc}

\newcommand\figwidth{0.8}
\newcommand{\msub}[1]{\ensuremath _{\mbox{\scriptsize #1}}}

\title{Free caloron gas in high temperature quenched QCD}

\ShortTitle{Free caloron gas in high temperature quenched QCD}

\author{Reka A.\ Vig \\
        University of Debrecen\\
        E-mail: \email{vig.reka@atomki.mta.hu}}

\author{\speaker{Tamas G.\ Kovacs} \\
        Institute for Nuclear Research, Debrecen \\
        E-mail: \email{kgt@atomki.mta.hu}}

\abstract{Across the finite temperature transition to the quark-gluon plasma,
  the QCD topological susceptibility decreases sharply. Thus in the high
  temperature phase the remaining topological objects (possibly calorons) form
  a weakly interacting dilute gas. The overlap Dirac operator, through its
  exact zero modes, allows one to measure the net topological charge. We show
  that separately the number of positively and negatively charged topological
  objects can also be extracted from the low end of the overlap Dirac
  spectrum. We find that slightly above the phase transition their number
  distributions are already consistent with an ideal gas of non-interacting
  topological objects.}

\FullConference{37th International Symposium on Lattice Field Theory -
  Lattice2019\\ 16-22 June 2019\\ Wuhan, China}

\begin{document}

\section{Introduction}

\begin{figure}
\begin{center}
 \includegraphics[width=0.40\columnwidth,keepaspectratio]{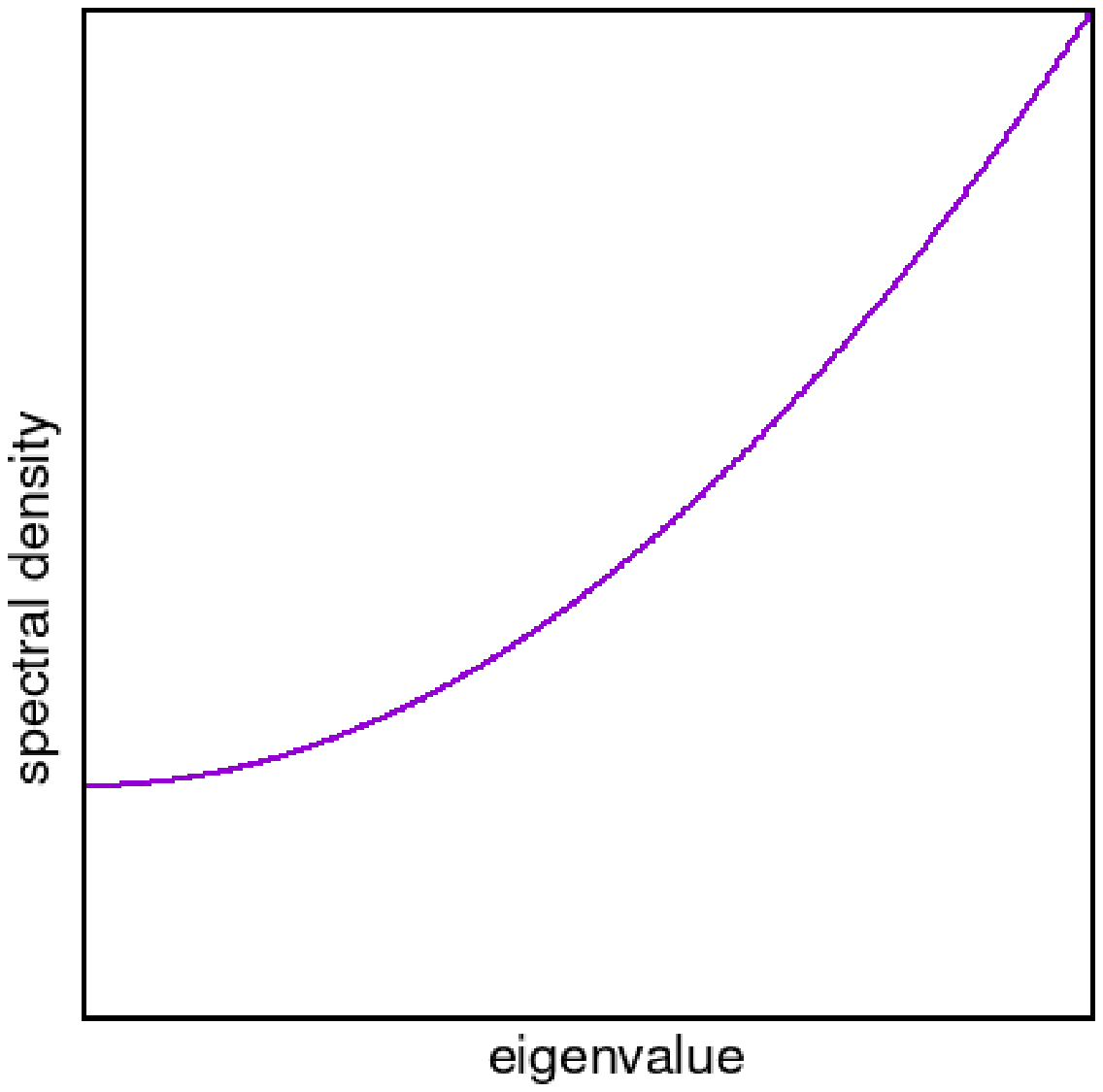}
    \hfill
 \includegraphics[width=0.40\columnwidth,keepaspectratio]{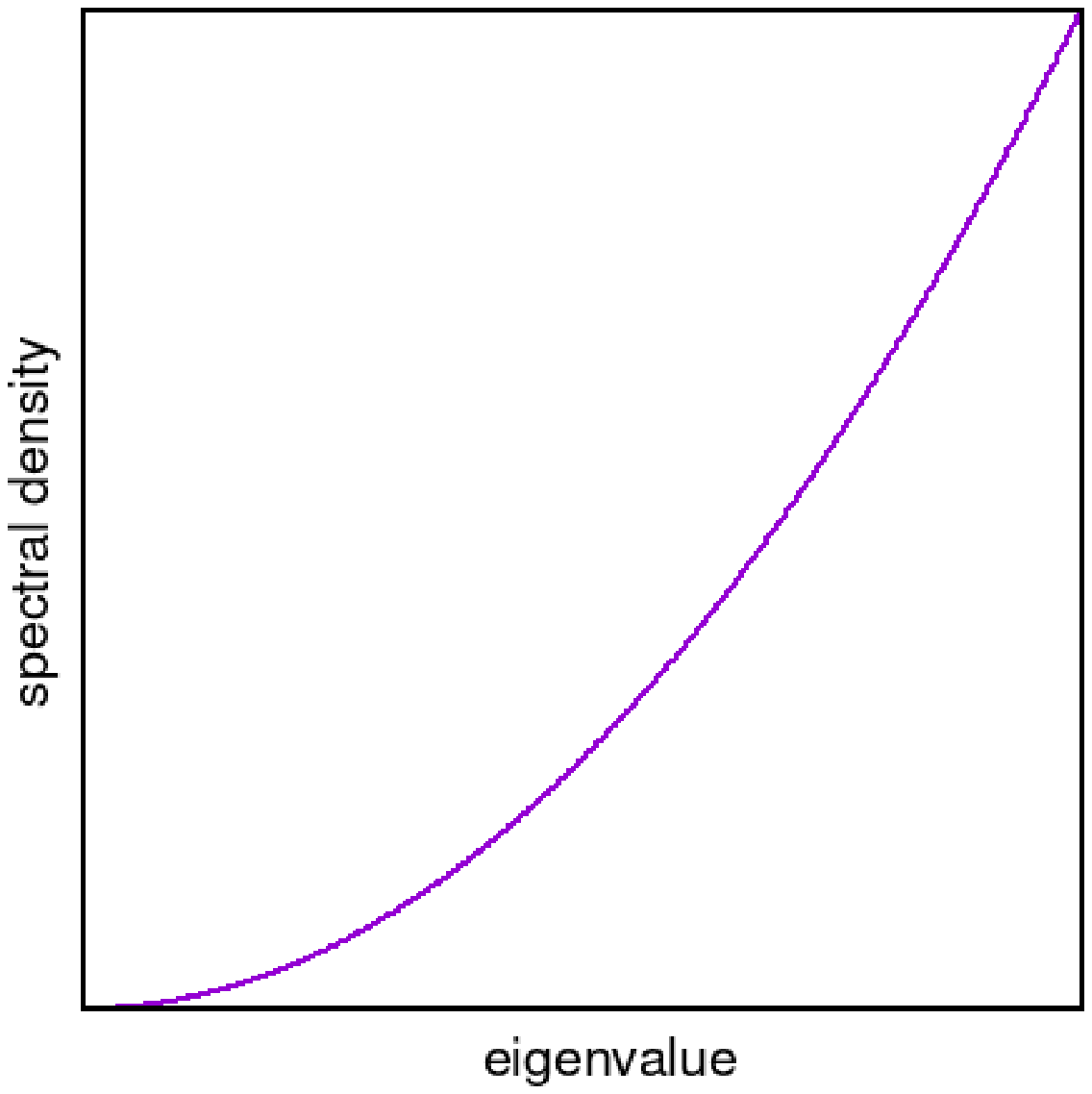}
\caption{\label{fig:chischem} Schematic depiction of the spectral density of
  the Dirac operator around zero virtuality at low temperature (left panel)
  and high temperature (right panel). According to the Banks-Casher relation
  the spectral density at zero is the order parameter of chiral symmetry. }
\end{center}
\end{figure}

It is well known that the finite temperature crossover of QCD is accompanied
by rapid changes in the Polyakov loop and the quark condensate, the order
parameters of confinement and chiral symmetry. In particular, as the
(approximate) chiral symmetry, spontaneously broken at zero temperature, gets
restored above the crossover, the quark condensate drops sharply. Since the
Banks-Casher \cite{Banks:1979yr} relation connects the quark condensate to the
spectral density of the Dirac operator around zero, the latter also falls
sharply as the system crosses into the high temperature quark-gluon plasma
phase. According to the traditional picture of chiral restoration, as shown in
the schematic plots of Fig.\ \ref{fig:chischem}, the spectral density at zero
becomes vanishingly small in the high temperature phase.

However, in reality the behavior of the spectral density across the transition
might be more complicated. Based on a study of the spectrum of the overlap
Dirac operator in quenched gauge backgrounds, it was already suggested a long
time ago that just above the transition temperature the spectral density at
zero does not vanish, but develops a ``spike'' \cite{Edwards:1999zm}. Light
dynamical fermions are expected to suppress low Dirac eigenvalues, as those
eigenvalues contribute small factors to the Dirac determinant. Thus the excess
of small eigenvalues and the resulting spike in the spectrum was initially
thought to be a quenched artifact. However, more recently it was suggested that
even light quarks cannot completely eliminate the spectral spike, at least as
long as the quarks are not exactly massless
\cite{Alexandru:2015fxa,Dick:2015twa}. Already the authors of
\cite{Edwards:1999zm} raised the possibility that this excess of low Dirac
eigenvalues may be connected to topological objects in the gauge field, and
the distribution of the number of eigenvalues in the spike supported this
assumption.

In the present paper we study this question in detail using large ensembles of
quenched gauge field configurations in the high temperature phase and the
overlap Dirac operator. Our main result is that around zero virtuality in the
Dirac spectrum there is a clearly separated region that is very likely to be
connected to topological objects in the gauge field. Above $T_c$ the Dirac
spectrum can thus be used to identify not only the net topological charge
through zero modes, but also the total number of topological objects.  This
enables us to examine their interactions, and in this quenched study we do not
find any trace of interactions among topological objects, to a very good
approximation they can be described as a dilute non-interacting gas.

\section{Topology and the zero mode zone}

In this section we briefly discuss how the lowest part of the Dirac spectrum
can be connected to topological fluctuations of the gauge field. It is an old
idea that the nonzero density of Dirac eigenvalues around zero that is needed
for the spontaneous breaking of chiral symmetry, is provided by mixing
approximate zero modes of a densely packed medium of instantons and
antiinstantons, called the instanton liquid (see \cite{Schafer:1996wv} for a
review). The mixing of topological zero modes can be easily understood by
considering an instanton antiinstanton pair separated by a distance much
larger than their size. If the separation were infinite, the Dirac operator
would have two opposite chirality zero modes. However, an arbitrarily small
perturbation, due to the two objects being a finite distance apart, will split
the degeneracy of the zero modes, resulting in two complex conjugate
eigenvalues of very small magnitude. Their magnitude, i.e.\ their splitting
from zero is mainly controlled by the distance of the topological objects; the
closer they are, the larger the perturbation of the zero modes and the larger
the splitting will be.

Unfortunately, in QCD at zero temperature, even if there are instantons, they
are too closely packed for this picture to be applicable in a straightforward
way. In particular, it is not possible to identify those small non-zero
eigenvalues in the spectrum that are connected to topology. This is because
the spectral density is monotonically rising everywhere, and the lowest part
of the spectrum, termed the zero mode zone (ZMZ) is smoothly connected to the
bulk of the spectrum. The situation, however, changes drastically as the
system crosses over into the high temperature, chirally restored phase. As the
temperature increases and the (Euclidean) temporal size of the system shrinks,
bigger instantons are squeezed out and fluctuations of the topological charge
fall sharply, the instanton liquid turns into a dilute gas of finite
temperature topological objects, calorons. As the typical distance between
topological objects increases, the splitting of the approximate zero modes
becomes ever smaller and these small Dirac eigenvalues are expected to be
present in the spectrum with a finite volume density, their density
corresponding to that of the topological objects surviving at the given
temperature. The question is whether these small eigenvalues of topological
origin can be identified in the spectrum and separated from the bulk modes,
the ones not related to topology.

\section{Identification of the ZMZ above $T_c$}

\begin{figure}
\begin{center}
 \includegraphics[width=\figwidth\columnwidth,keepaspectratio]{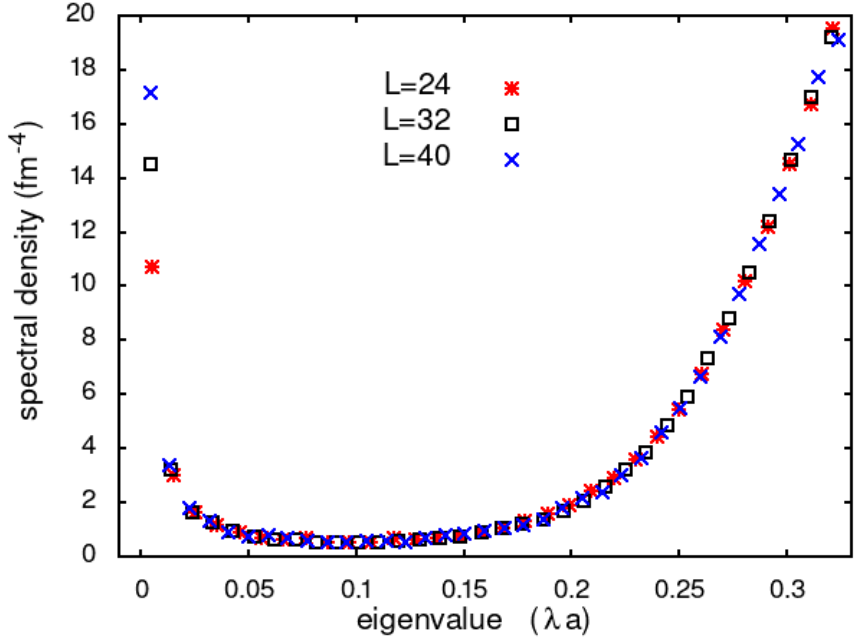}
    \hfill
\caption{\label{fig:spd} The spectral density of the overlap Dirac operator in
quenched gauge field backgrounds at $T=1.04T_c$. The temporal extension of the
lattice is $N_t=6$ and the different symbols correspond to spatial linear
extensions $L=24, 32, 40$.}
\end{center}
\end{figure}

In this section, based on numerical evidence we argue that the accumulation of
small Dirac eigenvalues above $T_c$ can indeed be connected to topology, and
we also demonstrate that the ZMZ can be clearly identified in the spectrum of
the overlap Dirac operator. In Fig.\ \ref{fig:spd} we show the spectral
density of the overlap operator on ensembles of quenched $SU(3)$ gauge field
configurations on lattices of temporal size $N_t=6$ and different spatial
volumes. The temperature is $T=1.04T_c$, just above the transition, which in
the quenched case, unlike in QCD with physical quarks, is a genuine phase
transition. The most striking feature of the spectral density is that instead
of smoothly and monotonically vanishing at $\lambda=0$, it has a spike there.
We emphasize that the spike seen in the figure is solely due to near zero
modes, as the exact zero modes have been removed from the spectral density.

It is tempting to identify the spike with the zero mode zone, i.e.\ the mixing
near zero modes connected to fluctuations of the gauge field topological
charge. How could this assumption be checked? One possibility is to check
whether the average number of eigenmodes in the spike and their distribution
are consistent with the fluctuations of the net topological charge, as given by
the topological susceptibility.

Since we computed overlap spectra on these configurations, we have the number
of exact zero modes, i.e.\ $n\msub{i}-n\msub{a}$, the difference of the number
of instantons and antiinstantons\footnote{Strictly speaking, at finite
  temperature the exact solutions of the Euclidean equations of motion for the
  gauge field are calorons. Nevertheless, we will freely use the words
  instanton and caloron for approximate solutions at high temperature.},
configuration by configuration. This in turn determines the topological charge
$Q$ and the topological susceptibility as
\begin{equation}
  \chi = \frac{1}{V} \langle Q^2 \rangle =
         \frac{1}{V} \langle (n\msub{i}-n\msub{a})^2 \rangle,
\end{equation}
where $\langle.\rangle$ denotes averaging with the path integral measure and
$V$ is the space-time volume. If the eigenmodes in the spike of the spectral
density are the ones corresponding to topological charge, then their number
$n\msub{spike}$ together with that of the exact zero modes is equal to the
total number of topological objects, i.e.\
\begin{equation}
  n\msub{i}+n\msub{a} = n\msub{spike} + |Q|.
\end{equation}

These numbers and their distributions are in principle independent and
complicated variables, determined by the dynamics of the system. However, they
are all connected in a simple way if topological objects occur statistically
independently. Fortunately, at high temperature topological fluctuation are
suppressed, calorons are expected to form a dilute gas and the interaction
among topological objects becomes smaller. If the interaction is completely
neglected then all the statistical properties of the distribution of
topological objects are encoded in a single parameter, for example the
topological susceptibility $\chi$. In this case it is easy to show that the
distribution of the number of instantons and antiinstantons in the gauge
configurations are given by two independent and identical Poisson
distributions with expectation $\frac{1}{2} V\chi$.  It is also not hard to
show that the probabilities of the different charge sectors $Q$ are given by
\begin{equation}
  P\msub{Q} = \mbox{e}^{-\chi V} I\msub{Q}(\chi V),
    \label{eq:qdist}
\end{equation}
where $I\msub{Q}$ are the modified Bessel functions of order $Q$. 

\begin{figure}
\begin{center}
 \includegraphics[width=\figwidth\columnwidth,keepaspectratio]{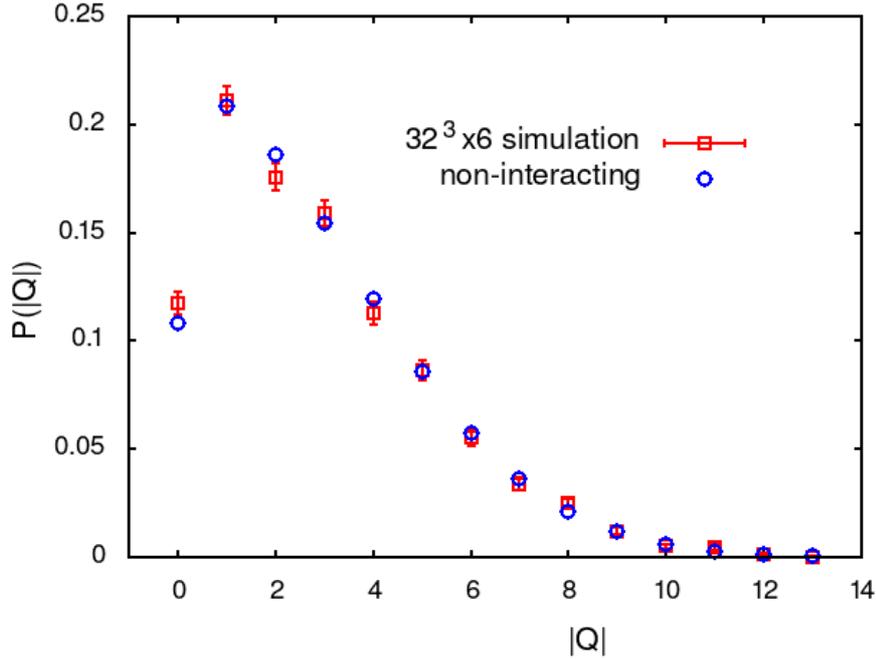}
    \hfill
\caption{\label{fig:qdist} The probability of the different topological charge
  sectors obtained from overlap zero modes on a set of 3800 $32^3 \! \times \!
  6$ lattice configurations at $T=1.04T\msub{c}$ (squares). Due to the
  expected symmetry, the probabilities of the charge sectors $Q$ and $-Q$ were
  added. The circles show the expected probabilities in a free instanton gas
  with the same topological susceptibility.}
\end{center}
\end{figure}

To test whether the caloron gas is non-interacting, in Fig.\ \ref{fig:qdist}
we compare the numerically obtained charge distribution with the distribution
of Eq.\ (\ref{eq:qdist}) expected for free calorons. The parameter $\chi$ of
the latter distribution was set equal to the numerically obtained topological
susceptibility, so the comparison effectively amounts to a one-parameter
fit. Surprisingly, even at this relatively low temperature, $T=1.04T\msub{c}$,
no significant discrepancy can be observed between the non-interacting
instanton model and the lattice data.

Encouraged by the good agreement of the lattice data and the free instanton
gas description, we can extend our study to the full zero mode zone, including
also the near zero modes. Again, based on the free instanton gas, it is easy
to see that the distribution of $n\msub{i}+n\msub{a}$ is Poisson, with
expectation $V\chi$. To be able to check this on the lattice data, we have to
fix an additional parameter, a cut in the spectrum, $\lambda\msub{zmz}$, below
which eigenvalues are considered to belong to the zero mode zone. The most
natural way to fix $\lambda\msub{zmz}$ is to require that
\begin{equation}
  \langle n\msub{i} + n\msub{a} \rangle = V \chi
    \label{eq:topobj}
\end{equation}
on the given ensemble of gauge configurations, where $\chi$ is again the
topological susceptibility obtained by counting zero modes. In a free instanton
gas eq.\ (\ref{eq:topobj}) holds exactly and for a lattice ensemble of
configurations, by definition $\langle n\msub{i} + n\msub{a} \rangle$ is the
average number of eigenvalues with $|\lambda|<\lambda\msub{zmz}$, including
exact zero modes.

\begin{figure}
\begin{center}
 \includegraphics[width=\figwidth\columnwidth,keepaspectratio]{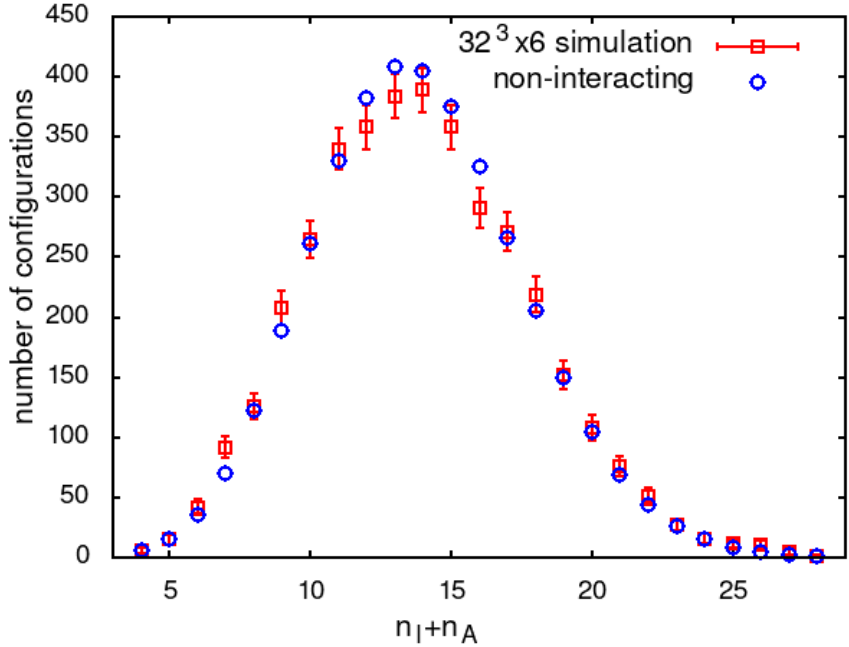}
    \hfill
\caption{\label{fig:iadist} The occurrence of different numbers of topological
  objects (instantons plus antiinstantons) obtained from overlap near zero
  modes on a set of 3800 $32^3 \!  \times \! 6$ lattice configurations at
  $T=1.04T\msub{c}$ (squares) compared to the numbers expected for an ideal
  gas of calorons (circles). The cut for the counted near zero modes was set
  to $\lambda a = 0.092$, consistently with the susceptibility, as explained
  in the text.}
\end{center}
\end{figure}

Fixing $\lambda\msub{zmz}$ in this way, we compare the numerically obtained
and the expected distribution in Fig.\ \ref{fig:iadist}, and we find good
agreement again. The distributions for the two other spatial lattice volumes,
$24^3$ and $40^3$ look similar and neither of them shows any significant
deviation from the distributions expected in a free caloron gas. The values of
$\lambda\msub{zmz}$ determined on the three ensembles are consistent with one
another and their combined average is $\lambda\msub{zmz}a =0.09\pm 0.01$. As a
further consistency check we can see in Fig.\ \ref{fig:spd} that this point in
the spectrum roughly coincides with the minimum of the spectral density
between the spike at zero and the bulk. This indicates that the modes of
topological origin can indeed be separated from the bulk of the spectrum. We
also expect this separation to become more pronounced in the continuum limit
and also at higher temperatures.

Another interesting feature of the spectral density in Fig.\ \ref{fig:spd} is
that for very small values of $\lambda $ (most conspicuously in the first bin)
there are unusually large corrections to the proper volume scaling of the
spectral density. However, as a more detailed analysis reveals, the total
number of topological modes scales properly with the volume. The discrepancy
in the figure is caused by the fact that the topological modes corresponding
to the net charge occur exactly at zero, but the density of these zero modes
vanishes with $1/\sqrt{V}$ and they will not contribute to the spectral
density in the thermodynamic limit. However, at these volumes their
contribution is still seen, resulting also in significantly smaller spectral
densities at the smallest $\lambda$ values.

\section{Conclusions}

In the present paper we gave strong indications that in quenched QCD already a
bit above $T\msub{c}$ the topology related near zero modes of the overlap
Dirac operator can be separated from the bulk of the spectrum. Counting these
modes, as well as the exact zero modes configuration by configuration, we
found the resulting distributions to be consistent with the ones expected for
a free caloron gas. The present study was based on the quenched approximation,
where the quark determinant is neglected. However, the Dirac determinant with
light quarks is expected to be sensitive to small Dirac eigenvalues and
through them also to topology which in turn could result in interactions among
topological objects. Therefore, the most interesting extension of this work
would be to repeat the study with the inclusion of light dynamical
quarks. Unfortunately, the presence or absence of quark induced interactions
among topological objects could crucially depend on the chiral properties of
the Dirac operator, i.e.\ how well it can resolve zero and near zero modes. It
would also be interesting to see how these topological modes are related to
the possible new phases of QCD discussed in Ref.\ \cite{Alexandru:2019gdm}.


\begin{thebibliography}{99}

\bibitem{Banks:1979yr} 
  T.~Banks and A.~Casher,
  Nucl.\ Phys.\ B {\bf 169}, 103 (1980).
  doi:10.1016/0550-3213(80)90255-2.


\bibitem{Edwards:1999zm} 
  R.~G.~Edwards, U.~M.~Heller, J.~E.~Kiskis and R.~Narayanan,
  Phys.\ Rev.\ D {\bf 61}, 074504 (2000)
  doi:10.1103/PhysRevD.61.074504
  [hep-lat/9910041].

  
\bibitem{Alexandru:2015fxa} 
  A.~Alexandru and I.~Horváth,
  Phys.\ Rev.\ D {\bf 92}, no. 4, 045038 (2015)
  doi:10.1103/PhysRevD.92.045038
  [arXiv:1502.07732 [hep-lat]].

\bibitem{Dick:2015twa} 
  V.~Dick, F.~Karsch, E.~Laermann, S.~Mukherjee and S.~Sharma,
  Phys.\ Rev.\ D {\bf 91}, no. 9, 094504 (2015)
  doi:10.1103/PhysRevD.91.094504
  [arXiv:1502.06190 [hep-lat]].

\bibitem{Schafer:1996wv} 
  T.~Schäfer and E.~V.~Shuryak,
  Rev.\ Mod.\ Phys.\  {\bf 70}, 323 (1998)
  doi:10.1103/RevModPhys.70.323
  [hep-ph/9610451].

  
\bibitem{Alexandru:2019gdm} 
  A.~Alexandru and I.~Horvath,
  arXiv:1906.08047 [hep-lat]; see also I.\ Horváth's contribution to this
  conference. 

 
  
\end{thebibliography}
\end{document}